\documentclass[aps,prc,groupedaddress,twocolumn,showpacs,amsmath,amssymb]{revtex4}
\usepackage{graphics}
\usepackage{dcolumn}
\usepackage{longtable}

\begin{document}

\title{Role of consistent parameter sets in an assessment of the alpha-particle optical potential below the Coulomb barrier}
\author{V.~Avrigeanu} \email{vlad.avrigeanu@nipne.ro}
\author{M.~Avrigeanu}
\affiliation{Horia Hulubei National Institute for Physics and Nuclear Engineering, P.O. Box MG-6, 077125 Bucharest-Magurele, Romania}

\begin{abstract}
\noindent
{\bf Background:} 
Further studies of high-precision measurements of $\alpha$-induced reaction data below the Coulomb barrier have still raised questions about the $\alpha$-particle optical model potential (OMP) within various mass ranges, i.e. for $^{64}$Zn, $^{108}$Cd, $^{113,115}$In, $^{121,123}$Sb, and $^{191,193}$Ir target nuclei.  

\smallskip
\noindent
{\bf Purpose:} 
The accuracy as well as eventual uncertainties and/or systematic errors of using a previous optical potential are much better considered by analysis of such accurate data. 

\smallskip
\noindent
{\bf Method:} 
Statistical model (SM) calculations of the $(\alpha,x)$ reaction cross sections have been carried out using model parameters which were previously obtained by analyzing independent data, particularly $\gamma$-ray strength functions, and taking into account their uncertainties and questions of extrapolation for nuclei without similar data. 

\smallskip
\noindent
{\bf Results:} 
Consistent description of the recent $\alpha$-induced reaction data is provided by the above-mentioned optical potential with no empirical rescaling factors of either its own parameters or the $\gamma$ and/or nucleon widths. Effects of still uncertain SM parameters on calculated $\alpha$-induced reaction cross sections and conclusions on the $\alpha$-OMP can be now discussed due to unprecedented precision of the new data.

\smallskip
\noindent
{\bf Conclusions}: 
The $\alpha$-particle optical potential has been confirmed at incident energies below the Coulomb barrier using  statistical-model parameters validated through a former analysis of independent data. 
\end{abstract}


\maketitle

\section{INTRODUCTION}
More recent high-precision measurements of $\alpha$-particle scattering and induced reaction data below the Coulomb barrier $B$ \cite{ao16,pm17,ps16,ggk16,ggk18,zk18,ts18} provide further opportunities to check a previous optical-model potential (OMP) of $\alpha$-particles on nuclei within the mass number range 45$\leq$$A$$\leq$209 \cite{va14}. It may thus complement the discussion of data provided in the meantime \cite{va16}. 

Actually, a semi-microscopic double-folding model (DFM) real part and the dispersive contribution of a phenomenological energy-dependent imaginary-potential were firstly involved within an analysis of $\alpha$-particle elastic-scattering angular distributions above $B$ \cite{va14,ma03,ma09}. Subsequently, the phenomenological real potential \cite{va14} was established using the same data basis. 
Then, the Hauser-Feshbach statistical model (SM) analysis of $\alpha$-induced reaction cross sections proved the particular energy dependence of the surface imaginary potential at incident energies below $B$ \cite{va14,ma09,ma09a,ma10a,ma10}. 
The changes corresponding to the correction due to the dispersive relations with an integral over all incident energies, in the real part of the semi-microscopic potential, as well as the phenomenological OMP \cite{va14} are within uncertainties of the parameter values in the rest of the energy range \cite{ma09}. Particular comments have concerned the decreasing side of the volume integral per nucleon of the imaginary surface potential (Fig. 9 of Ref. \cite{va14}) which is constrained by the elastic-scattering data while the increasing one could be determined by means of the $\alpha$-induced reaction data analysis.

A key point of these analyzes \cite{va14,ma09,ma09a,ma10a,ma10} has been the use of no empirical rescaling factors of the $\gamma$ and/or neutron widths but consistent parameter sets. 
The description of all available $\alpha$-induced reaction data, of equal interest for astrophysics and nuclear technology, have thus been obtained.
On the other hand, a former OMP \cite{va94} concerned only the $\alpha$-particle emission in neutron-induced reactions, with distinct predictions from potentials for incident $\alpha$ particles \cite{lmf66}. 
Thus, this OMP \cite{va94} was not an earlier version of the above-mentioned ones \cite{va14,ma03,ma09}, while the question on different OMPs for incident and emitted $\alpha$ particles \cite{va94,ma06,va15,va17} is still unanswered.

As the above-mentioned new studies have raised $\alpha$-OMP questions yet open within various mass ranges, similar analyzes of their data become mandatory for the assessment of potential \cite{va14}, before being taken into consideration for the tentative account of $\alpha$-emission as well. 
Thus, a reasonable description of additional $\alpha$-scattering and induced reaction cross sections on $^{64}$Zn at low energies \cite{ao16} was considered to be provided by several $\alpha$-particle OMPs including \cite{va14}. But then a further $\chi^2$-based analysis concluded that neither a better model for calculation of the $^{64}$Zn+$\alpha$ reaction cross sections nor better parametrizations of the SM ingredients are available \cite{pm17}. 
The need for further improvement of SM calculations and particularly the $\alpha$-nucleus potential was thus deemed necessary.

Moreover, precise cross sections of the $(\alpha,\gamma)$ and $(\alpha,n)$ reactions on $^{108}$Cd, measured for first time close to the astrophysically relevant energies, have been found to provide further support to investigations of the real part of $\alpha$-OMP in order to improve the understanding of reactions involving $\alpha$ particles \cite{ps16}. 
Simultaneous measurement of the $(\alpha,\gamma)$ and $(\alpha,n)$ reactions on $^{115}$In, in addition to high-precision elastic scattering \cite{ggk16} and together with a best-fit combination of all SM parameters, also concluded that further improvements of the $\alpha$-nucleus potential are still required for a global description of elastic scattering and $\alpha$-induced reactions in a wide range of masses and energies \cite{ggk18}. 
It was particularly surprising to find a significant underestimation of isomeric $(\alpha,n)$ and $(\alpha,\gamma)$ cross sections by the OMP \cite{va14} at once with an excellent description of the elastic scattering data, while the largest deviation from elastic scattering angular distributions is shown by a potential with the best description of the $(\alpha,x)$ data. 
It was thus concluded that further efforts are needed to establish an OMP which simultaneously describes $\alpha$-particle elastic scattering and reaction data \cite{ggk18}.

On the other hand, a first measurement on $^{121}$Sb close to the astrophysically relevant energy range  \cite{zk18} provided further support to the conclusion that experimental $(\alpha,\gamma)$ data, where they exist, are often strongly overestimated by SM calculations. 
A similar overestimation of $(\alpha,\gamma)$ measurements with a previously unprecedented sensitivity on $^{191,193}$Ir has also been obtained within a SM analysis which reproduced well the $(\alpha,n)$ data \cite{ts18}.

The aim of the present work is to analyze, in addition to \cite{va16}, the SM results provided by the previous $\alpha$-particle optical potential \cite{va14} in the case of the new data \cite{ao16,ps16,pm17,ts18,zk18,ggk18}, as well as their uncertainties and possible systematic errors. 
The use of consistent input parameters established or validated by analyzing various independent data (e.g., Ref. \cite{eda80}) constitutes the essential difference between our analysis and the above-mentioned data fit using a range of SM global input parameters. 

While detailed presentation of  model parameters was given in Refs. \cite{va14,va15,va16,ma12}, latest particular parameter values are given in Sec.~\ref{SMcalc} of this work. 
The SM results obtained using the OMP of Ref. \cite{va14} are then compared with the above-mentioned measured cross sections \cite{ao16,ps16,pm17,ts18,zk18,ggk18} in Sec.~\ref{Res}, followed by conclusions in Sec.~\ref{Conc}.
Preliminary results were described elsewhere  \cite{cssp18}.

\section{Statistical model parameters} \label{SMcalc}

SM calculations discussed in the following section were carried out within a local approach using an updated version of the computer code STAPRE-H95 \cite{ma95}, with $\sim$0.1-0.3 MeV equidistant binning for the excitation energy grid.
The direct-interaction (DI) distorted-wave Born approximation (DWBA) method and a local version of the code DWUCK4 \cite{pdk84} were also used for calculation of the collective inelastic-scattering cross sections using the corresponding deformation parameters \cite{sr01,tk02} of the first 2$^+$ and 3$^-$ collective states.  The collective form of the Coulomb excitation (CE) has been considered in the usual way \cite{pdk84}, while the comments on CE effects given in Sec. II.A of Refs. \cite{va14,va16} apply here as well. 
The calculated DI cross sections are then involved for the subsequent decrease of the total-reaction cross sections $\sigma_R$ that enter SM calculations. 
Typical DI inelastic-scattering cross sections, e.g. for $^{64}$Zn target nucleus, grow up from $\sim$11\% to $\sim$18\% of $\sigma_R$ for $\alpha$-particle energies from 6.6 to 8 MeV, and then decrease to $\sim$3\% at the energy of $\sim$16 MeV.

The consistent set of nucleon and $\gamma$-ray transmission coefficients, and back-shifted Fermi gas (BSFG) \cite{hv88} nuclear level densities (NLDs) were established or validated using independently measured data as the neutron total cross sections and $(p,n)$ reaction cross sections \cite{exfor}, $\gamma$-ray strength functions \cite{jk17,OCL} and $(p,\gamma)$ reaction cross sections \cite{exfor}, and low-lying levels \cite{ensdf} and resonance data \cite{hv88,ripl3}, respectively. 
The details in addition to the ones given formerly \cite{va14,va15,va16,ma12} as well as particular parameter values are mentioned below in order to provide the reader with all main details and assumptions of the present analysis.

The reaction cross sections calculated within this work are also compared with the content of the evaluated data library TENDL-2017 \cite{TENDL} provided by using the code TALYS-1.9 \cite{TALYS}, for an overall excitation function survey. 

\begingroup
\squeezetable
\begin{table*}  
\caption{\label{densp} Low-lying levels number $N_d$ up to excitation energy $E^*_d$ \protect\cite{ensdf} used in cross-section SM calculations, 
the low-lying levels and $s$-wave nucleon-resonance spacings $D_0^{\it exp}$ (with uncertainties given between parentheses, in units of the last digit) in the energy range $\Delta$$E$ above the separation energy $S$, for the target-nucleus g.s. spin $I_0$, fitted to obtain the BSFG level-density parameter $a$ and g.s. shift $\Delta$, for the given ratio $I/I_r$ for excitation energies between g.s. and $S$, the average $s$-wave radiation widths $\Gamma_{\gamma}$, either measured \cite{ripl3} or based on systematics (given between square brackets), and corresponding to the SLO, GLO, and EGLO models, with the parameter $T_f$ of the EGLO model obtained by description of the RSF data \cite{OCL}.
}
\begin{ruledtabular}
\begin{tabular}{cccccccccccccccc}
Nucleus&$N_d$&$E^*_d$&\multicolumn{6}{c}{Fitted low-lying levels and nucleon-resonance data}&
				                                  $a$&$I/I_r$&$\Delta$&$T_f$&\multicolumn{3}{c}{$\Gamma_{\gamma}$}\\ 
		  \cline{4-9}                                                                          \cline{14-16}
 & & &$N_d$&$E^*_d$&$S+\frac{\Delta E}{2}$&$I_0$&$D_0^{\it exp}$&$\Gamma_{\gamma}$& & & & & SLO& GLO& EGLO\\ 
          &  &(MeV)&  &(MeV)& (MeV)&   & (keV) & (meV) &(MeV$^{-1}$)&      & (MeV)&(MeV)&(meV)&(meV)&(meV)\\ 
\hline
$^{67}$Ga &28&2.282&28&2.282& 8.420& 0 &2.5(2))$^a$&      &8.20(8)&0.5-0.75&-0.55 & \\ 
$^{67}$Ge &21&1.747&21&1.747&      &   &       &          &8.05(8)&0.5-0.75&-0.95(16) & \\ 
$^{68}$Ge &16&3.087&16&3.087&12.392&1/2&       &[550(200)]& 8.3(3)&0.5-0.75& 0.72 & 0.5 & 1890 & 1700& 575\\ 
$^{112}$Sn&23&2.986&23&2.986&10.786&7/2&      &[140(40)]&13.85(40)&0.5-0.75& 1.34 & 0.46 & 311 & 215 & 106\\ 
$^{117}$Sb&17&1.536&18&1.624& 9.889& 3 &       &[140(50)]& 14.1(4)&0.5-0.75& 0.10 & 0.46 & 477 & 344 & 157\\ 
$^{119}$Sb&23&1.676&23&1.676& 9.549& 1 &       &[140(50)]& 14.4(4)&0.5-0.75& 0.08 & 0.46 & 433 & 365 & 140\\
$^{124}$I &50&0.725&51&0.748&      &   &       &         & 15.5(6)&0.5-0.75&-1.10 \\
$^{125}$I &31&1.392&31&1.392& 9.543& 2 &       &[140(50)]& 14.6(6)&0.5-0.75&-0.32 & 0.60(14)&419&318 & 180\\
$^{126}$I &26&0.410&30&0.458&      &   &       &         & 14.8(6)&0.5-0.75&-1.36 \\
$^{127}$I &33&1.480&33&1.480&      &   &       &         & 14.0(6)&0.5-0.75&-0.35 \\  
$^{195}$Au&36&1.443&36&1.443& 8.426& 1 &       & [128(6)]& 18.8(4)&    1   &-0.12 & 0.15 & 390 & 330 & 121\\ 
$^{198}$Au&28&0.549&28&0.549& 6.515&3/2&0.0155(8)$^b$&128(6)&17.50(9)&  1  &-1.12 & 0.15 & 380 & 340 & 128\\ 
\end{tabular}	 
\end{ruledtabular}
\begin{flushleft}
$^a$Reference \cite{hv88}.\\
$^b$Reference \cite{ripl3}.\\
\end{flushleft}
\end{table*}
\endgroup

\subsection{Nuclear level densities} \label{NLD}

The BSFG parameters used to obtain the present SM results, which are either updated or not already provided in Refs. \cite{va14,va15,va16,ma12}, are given in Table~\ref{densp}. They follow the low-lying level numbers and corresponding excitation energies \protect\cite{ensdf} used in the SM calculations (the 2nd and 3rd columns) as well as those fitted at once with the available nucleon-resonance data \cite{hv88,ripl3} to obtain these parameters. The level-density parameter $a$ and ground state (g.s.) shift $\Delta$ were generally obtained with a spin cutoff factor corresponding to a variable moment of inertia $I$, between half of the rigid-body value $I_r$ at  g.s., 0.75$I_r$ at the separation energy $S$, and the full $I_r$ value at the excitation energy of 15 MeV, with a reduced radius $r_0$=1.25 fm \cite{va02}. The only different case is that of Au isotopes, for which there is a definite proof for a constant $I_r$ value \cite{ma12}. 
 
The fit of the error-bar limits of $D_0^{\it exp}$ data has also been used to provide limits of the fitted $a$-parameters. Moreover, these limits are used within SM calculations to illustrate the NLD effects on the calculated cross-section uncertainty bands (Sec.~\ref{Res}). 

On the other hand, the smooth-curve method \cite{chj77} was applied for nuclei without resonance data, using average $a$-values of the neighboring nuclei with resonance data, to obtain only the $\Delta$ values by fit of the low-lying discrete levels. 
The uncertainties of these averaged $a$-values, following the spread of the fitted $a$ parameters, are also given in Table~\ref{densp}.
These uncertainties are obviously larger than those of the $a$-values obtained by fit of $D_0^{\it exp}$. Thus, use of their limits in SM calculations leads to increased NLD effects on calculated cross-section uncertainty bands; the same $\Delta$ values have been used within this uncertainty analysis, to take into account an usual uncertainty of 1--2 low-lying levels.

An additional question related to the $\Delta$-parameter value is taken into account in the particular case of $^{67}$Ge nucleus, for which only 4 excited levels at mid of the energy range $\sim$1.4--1.9 MeV are currently known.  
Thus, the NLD change due to this $\Delta$-uncertainty goes  over the above-mentioned usual ambiguity of 1--2 low-lying levels and is also considered within the accuracy discussion (Sec.~\ref{Zn}).

\subsection{$\gamma$-ray strength functions} \label{RSF}

The corresponding average $s$-wave radiation widths $\Gamma_{\gamma}$ \cite{ripl3} including the extrapolated values based on systematics and the $\Gamma_{\gamma}$ distinct $S$-dependence for even-even and odd-$A$ nuclei (e.g., Ref. \cite{hkt11}) are also provided in Table~\ref{densp}. 
They have been used together with earlier \cite{kn82,kn80,ripl2,be79,be80,kn83,as14,jk17} and more-recently \cite{OCL} measured radiative strength functions (RSF) data for validation of the $\gamma$-ray  transmission coefficients by using the former Lorentzian (SLO) \cite{pa62}, generalized Lorentzian (GLO) \cite{jk90}, and enhanced generalized Lorentzian (EGLO) \cite{jk93} models for the electric-dipole $\gamma$-ray strength functions. 
The recently-compiled \cite{RIPL3gamma} giant dipole resonance (GDR) line-shape parameters were used here. 
The constant nuclear temperature $T_f$ of the final states \cite{acl10}, which is particularly assumed within the EGLO model, is also given in Table~\ref{densp} as well as the calculated $\Gamma_{\gamma}$ values corresponding to the three electric-dipole RSF models. 

\begin{figure} 
\resizebox{1.0\columnwidth}{!}{\includegraphics{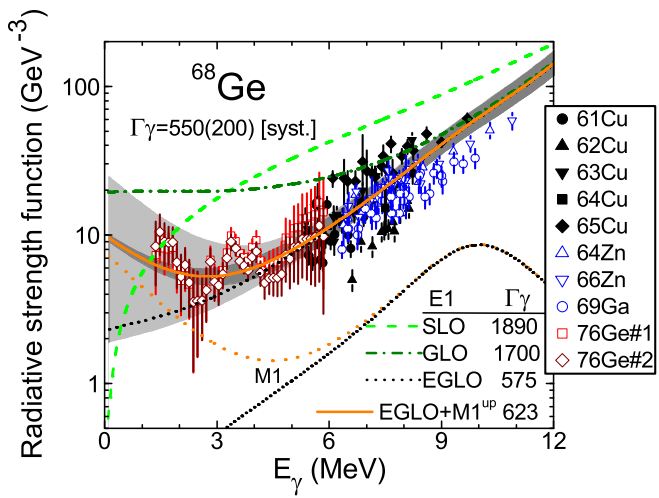}}
\caption{\label{Fig:RSF_GeF}(Color online) Comparison of the measured dipole $\gamma$-ray strength functions for $^{61,62,63,64,65}$Cu, $^{64,66}$Zn, $^{69}$Ga \cite{kn82,kn80,ripl2,be79,be80,kn83} and $^{76}$Ge nuclei \cite{as14}, and the sum of calculated $\gamma$-ray strength functions of the SLO model for $M$1 radiations (short-dotted curve), and $E$1-radiation models SLO (dashed curve), GLO (dash-dotted curve), EGLO (short-dashed curve), as well as the sum (solid curve) of the $M$1 component including the upbend to zero energy (dotted curve) and EGLO, for $^{68}$Ge nucleus. 
Uncertainties corresponding to those assumed (see text) for $E$1-radiation EGLO/GDR parameters (gray band) and, in addition, for the $M1$-radiation upbend (light-gray band) are shown too; 
$s$-wave average radiation widths $\Gamma_{\gamma}$ (in meV, also in Table~\ref{densp}) are deduced from systematics \cite{ripl3} or correspond to either $M1$ and each of above-mentioned $E$1 models, or the upbending $M1$ and EGLO models.} 
\end{figure}

Concerning the $M1$ radiation, the SLO model was mainly used alone, with the GDR parameters derived from photoabsorption data or the global parametrization \cite{ripl3} for the GDR energy and width, i.e. $E_0$=41/$A^{1/3}$ MeV and $\Gamma_0$=4 MeV. 
However, in the particular case of $^{68}$Ge nucleus, we also considered the exponential increase of this RSF at decreasing energies approaching zero, predicted by shell-model calculations following the experimental observation of a dipolar RSF upbend (\cite{sg18} and Refs. therein). 
Thus, the function $f_{up}(E_{\gamma})$=$C$exp$(-{\eta}E_{\gamma})$ has been added (e.g., Refs. \cite{as14,tr16}) to the SLO component of the $M$1 strength, with the average parameter values $C$=0.77$\times$10$^{-8}$ MeV$^{-3}$ and $\eta$=0.578 MeV$^{-1}$ found most recently for the $f_{5/2}pg_{9/2}$--shell nuclei \cite{jem18}. The slope thus obtained is not as steep as for $^{74,76}$Ge isotopes \cite{as14,tr16}, in agreement with the results of Midtb\o{}  {\it et al.} for $N$ mid-shell nuclei. Moreover, the trend of the total $E$1+$M$1 RSF in this case, corresponding to EGLO model and $M$1 upbend to zero energy (Fig.~\ref{Fig:RSF_GeF}), is compatible with measured data of neighboring nuclei especially within the main related uncertainties, as follows.

First, while the GDR parameters of $^{70}$Ge \cite{RIPL3gamma} were used also for $^{68}$Ge, we considered a systematical uncertainty of the EGLO form given by the difference between the GDR peak cross sections $\sigma_0$ for $^{70,72}$Ge nuclei \cite{RIPL3gamma}.
Thus, an electric-dipole strength uncertainty band that corresponds to $\sigma_0$=(88.4$\pm$16) mb is illustrated by the gray band in Fig.~\ref{Fig:RSF_GeF}.  
Second, limits of the insight of $M$1-radiation upbend-function $f_{up}(E_{\gamma})$ have been additionally assumed. 
Therefore, we have considered an upper limit $C$=3$\times$0.77$\times$10$^{-8}$ MeV$^{-3}$, using a multiplying factor previously used \cite{sg18}. A lower limit, given by the $M$1-upbend disregarding, corresponds to its yet general missing in $(\alpha,\gamma)$ cross-section calculations. The resulting total-uncertainty band shown in Fig.~\ref{Fig:RSF_GeF}, in order to facilitate comparison with various SM calculations, may rather overestimate the uncertainty of adopted RSF than underestimates it. 

The $s$-wave average radiation widths $\Gamma_{\gamma}$, either deduced from systematics \cite{ripl3} or corresponding to $M$1 and each of above-mentioned $E$1 functions, are also given in Table~\ref{densp} as well as in Figs.~\ref{Fig:RSF_GeF}--\ref{Fig:Ir19193ax} to provide an immediate comparison of RSF effects on both $\Gamma_{\gamma}$ and $(\alpha,\gamma)$ cross-section calculations.
A particular note concerns again $^{68}$Ge nucleus.
Actually, the $\Gamma_{\gamma}$ error bar that may be estimated in this case by using the measured data for even-even nuclei \cite{ripl3}, versus $S$, is rather large especially due to the greater $S$ value of $^{68}$Ge. 
However, while the EGLO model leads to a calculated $\Gamma_{\gamma}$ close to this inference, the GLO and SLO predictions are higher by more than 5 times its uncertainty. 
At the same time, the above-assumed uncertainty of the GDR parameters within the EGLO model provides $\Gamma_{\gamma}$ changes of $\sim$17 \% while the one including the $M$1 upbend is still only around 30 \%. 
The propagation of these RSF uncertainties on the calculated $(\alpha,\gamma)$ reaction cross sections is discussed next.

\section{Results and Discussion} \label{Res}
\subsection{$(\alpha,x)$ reactions on $^{64}$Zn} \label{Zn}

\begin{figure*} 
\resizebox{1.5\columnwidth}{!}{\includegraphics{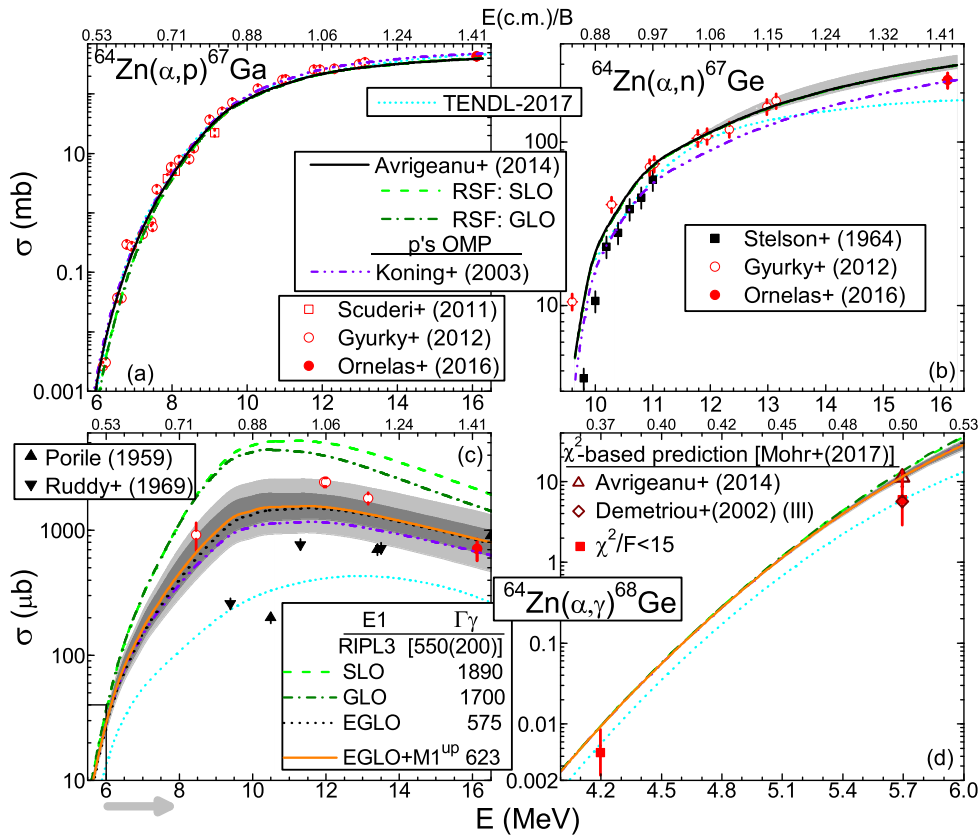}}
\caption{\label{Fig:Zn64ax}(Color online) Comparison of measured \cite{ao16,gg12,exfor}, evaluated within TENDL-2017 library \cite{TENDL} (short-dotted curves), and calculated cross sections of $\alpha$-induced reactions on $^{64}$Zn using the electric-dipole RSF models SLO (dashed curves), GLO (dash-dotted curves), EGLO (short-dashed curve) models along the SLO model for $M$1 radiation, as well as EGLO and the $M$1-radiation upbend (solid curves) for $^{68}$Ge nucleus, with the proton OMP of Ref. \cite{va15} and the alternate involvement of Ref. \cite{KD03} (dash-dot-dotted curves), versus laboratory energy of $\alpha$-particle (bottom) and corresponding ratio of the center-of-mass energy and Coulomb barrier $B$ \cite{wn80} (top); uncertainty bands correspond to (b) NLD parameters for $^{67}$Ge (light-gray band), and 
(c,d) those assumed for EGLO and $M$1-upbend RSFs (gray band) and, in addition, for NLD parameters of $^{68}$Ge nucleus (light-gray band); $\Gamma_{\gamma}$ values (in meV) are either based on data systematics \cite{ripl3} or corresponding to above-mentioned RSF models for $E$1 radiation; 
(d) comparison of presently calculated $(\alpha,\gamma)$ cross sections and $\chi^2$-based assessment of Mohr {\it et al.} \cite{pm17} for $\chi^2/F$$<$15 per data point (solid squares), and results \cite{pm17} using the $\alpha$-particle potentials of Refs. \cite{va14} (triangle) and \cite{pd02} (diamond).}
\end{figure*}

The use of $\alpha$-particle potential \cite{va14} provided already a suitable description  \cite{va15} of the $(\alpha,\gamma)$, $(\alpha,n)$, and $(\alpha,p)$ reaction data provided by Gy\"urky {\it et al.} \cite{gg12} for $^{64}$Zn at energies below $\sim$1.2$B$, as well as of the more recent $(p,\alpha)$ reaction cross sections for the same target nucleus \cite{gg14}.
The new high-precision data of Ornelas {\it et al.} \cite{ao16} are particularly worthwhile for the present work as they enlarge the incident-energy range for the three above-mentioned reactions (Fig.~\ref{Fig:Zn64ax}). This energy extension is particularly useful for the related $\alpha$-capture due to significant spreading of the earlier data [Fig.~\ref{Fig:Zn64ax}(c)] and existence of the newer data of Gy\"urky {\it et al.} at only three energies. These authors have also performed a careful extrapolation to low energies of these reaction cross sections within their newest analysis of the three reactions \cite{pm17}.

The results of our previous calculations \cite{va15} at the higher as well as lower $\alpha$-particle energies, using the same SM parameters and obviously the unchanged $\alpha$-particle potential \cite{va14}, are compared with the measured data in Fig.~\ref{Fig:Zn64ax}. 
The nucleon-OMP and RSF dependences shown formerly in Fig. 5 of Ref. \cite{va15} are also included in this figure.
The description of the new data \cite{ao16} is slightly different for the three reactions. 
First, one may note the good agreement for the major $(\alpha,p)$ reaction, with the changes due to different nucleon-OMP and RSF models within the limits of the experimental error bars [Fig.~\ref{Fig:Zn64ax}(a)].

Second, the measured $(\alpha,n)$ reaction cross section at the higher incident energy is described in the limit of 2$\sigma$ uncertainty [Fig.~\ref{Fig:Zn64ax}(b)] while the concurrence provided by the nucleon OMPs of Koning and Delaroche \cite{KD03} at lower energies is not confirmed by the proton OMP analysis \cite{va15} as well as the Gy\"urky {\it et al.} data \cite{gg12}. 
On the other hand, while use of the $a$-values for $^{67}$Ga corresponding to the fit of $D_0^{\it exp}$ error-bar limits (Table~\ref{densp}) leads to NLD effects within 3.5\% of the calculated $(\alpha,p)$ cross sections, a similar statement may concern $^{67}$Ge $a$-value but not its $\Delta$ parameter. Thus, a large ambiguity concerning the number of $^{67}$Ge low-lying levels to be fitted, between 17 and 22 in the energy range 1.432--1.901 MeV, provides limits of the fitted $\Delta$-value (Table~\ref{densp}) leading to the uncertainty band shown in Fig.~\ref{Fig:Zn64ax}(b).
This band has obviously risen only at the incident energies above the ones corresponding to population of the discrete levels. Nevertheless, it overestimates even more the newest data point at higher energy.

Third and most important, there is an entire agreement of the newly-measured $(\alpha,\gamma)$ cross section and calculated cross sections with the EGLO model for the electric-dipole RSF [Fig.~\ref{Fig:Zn64ax}(c)], in spite of the other data spreading and large variation of the results corresponding to the SLO and GLO models. 

We should also emphasize the increase of the calculated $(\alpha,\gamma)$ cross sections due to inclusion of the $M$1 upbend RSF component, of no more than $\sim$12\% around 8 MeV incident energy, and even $<$3\% below $\sim$5.5 MeV as well as above 12.5 MeV. However, this change should be compared with the calculated cross-section uncertainties [gray band in Fig.~\ref{Fig:Zn64ax}(c)] following the above-mentioned ones of the adopted RSF (shown at their turn by light-gray band in Fig.~\ref{Fig:RSF_GeF}), which is increasing from $\sim$2\%, at incident energies around 5 MeV, to $<$30\% at 8--9 MeV. 
On the other hand, the accuracy of these calculated cross sections really depends also on NLD parameters. 
In order to estimate their effects, we carried out SM calculations using the upper and lower limits of the level-density parameter $a$ of $^{68}$Ge, with the value (8.3$\pm$0.3) MeV$^{-1}$ obtained with the smooth-curve method \cite{chj77}. 
The corresponding uncertainty band is not particularly shown in Fig.~\ref{Fig:Zn64ax}(c)  because it overlaps with the one for the total RSF uncertainty. 
However, the uncertainty band corresponding to the sum of the above-mentioned RSF and NLD effects, with utmost change from 3\%, around the incident energy of 5 MeV, to 48\% at 16.5 MeV are displayed too. 
The calculated cross sections using the $E$1-radiation SLO and GLO models are only at incident energies $<$5 MeV inside this uncertainty band, while their increase over the EGLO results reaches then a factor of $\sim$3.

\begin{figure*} 
\resizebox{1.5\columnwidth}{!}{\includegraphics{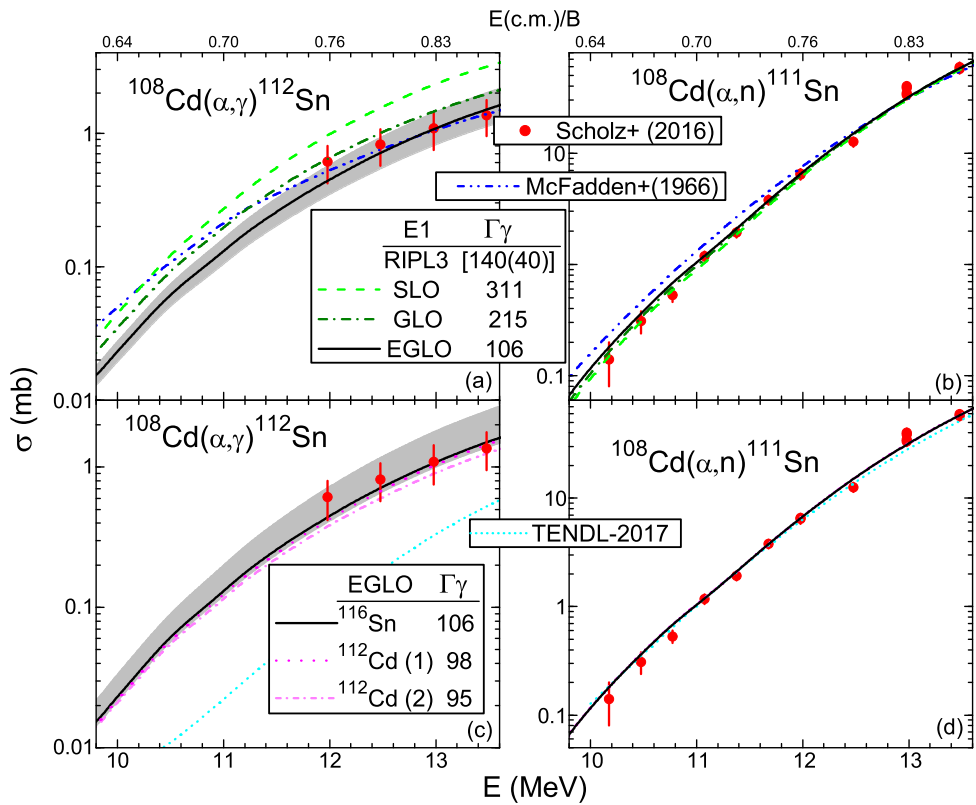}}
\caption{\label{Fig:Cd108ax}(Color online) As Fig.~\ref{Fig:Zn64ax} but for the target $^{108}$Cd and excited $^{112}$Sn nuclei \cite{ps16}, except for the use of the EGLO parameters of either $^{116}$Sn \cite{hkt11} (solid curves), or (c) $^{112}$Cd \cite{acl13} corresponding to $T_f$ parameter values of 0.4 MeV (dotted curve) and 0.37 MeV (short dash-dotted curve), the alternate use of (a,b) the $\alpha$-particle OMP of McFadden and Satchler \cite{lmf66} (dash-dot-dotted curves), and the uncertainties (light-gray bands) corresponding to limits of either (a) the level-density parameter $a$ for $^{112}$Sn or (c) $\Gamma_{\gamma}$ value (in meV) based on data systematics \cite{ripl3} (Table~\ref{densp}).}
\end{figure*}

The lowest-energy region of Fig.~\ref{Fig:Zn64ax}(c) is expanded in Fig.~\ref{Fig:Zn64ax}(d), in order to compare the present calculations with the results of the $\chi^2$-based assessment of the $(\alpha,\gamma)$ reaction cross sections at two incident energies of particular astrophysical interest, i.e., 3.95 and 5.36 MeV \cite{pm17}. 
This assessment corresponds to the best fit of recently measured data for the above-mentioned reactions using all available options of TALYS-1.8 for the $\alpha$-particle and nucleons OMPs, $\gamma$-ray strength functions, and nuclear level density. 
While the best fit shows $\chi^2/F$$\approx$7.7 per data point, a reasonable small $\chi^2/F$ was considered to satisfy the criterion $\chi^2/F$$<$15 per data point [the error bars in Fig.~\ref{Fig:Zn64ax}(d)]. 
The potential \cite{va14} led to an agreement close to the higher limit of this criterion. 

On the other hand, one may note that the smallest $\chi^2/F$$\approx$7.7 per point was derived using the SLO model at obvious variance with the independent analysis of RSF data \cite{va15}. 
At the same time it should be underlined that the lowest $\chi^2/F$ values for each reaction channel as well as for all of them correspond to different combinations of the above-mentioned four SM-parameter categories.
The case of TENDL-2017 evaluation, which shows the best agreement with the extrapolation below $\sim$0.5$B$ [Fig.~\ref{Fig:Zn64ax}(d)] while underestimating by at least a factor of 5 [Fig.~\ref{Fig:Zn64ax}(c)] the $(\alpha,\gamma)$ reaction cross sections \cite{ao16,gg12} recently measured  below $\sim$1.5$B$, is also open to discussion.

\subsection{$(\alpha,x)$ reactions on $^{108}$Cd} \label{Cd}

The analysis of precise cross sections of $(\alpha,\gamma)$ and $(\alpha,n)$ reactions on $^{108}$Cd measured for first time close to astrophysically relevant energies \cite{ps16} has completed a recent similar one for $^{106}$Cd including elastic-scattering angular distributions \cite{ao15}. 
Those data were already discussed \cite{va16} and proved to be well described by the optical potential \cite{va14} provided that suitable RSF are taken into account. 
However, the analysis of $^{108}$Cd data indicated that additional information about the RSF, for instance, are necessary to additionally test the $\alpha$-particle OMP \cite{ps16}. 

The previous analysis for $^{106}$Cd target nucleus \cite{va16} has been resumed for $^{108}$Cd with only one change, due to fact that the excited nucleus $^{112}$Sn is closer to $^{112}$Cd and $^{116}$Sn, with recent RSF data \cite{acl13,hkt11} already reviewed in Fig. 3 of \cite{va16}. 
Thus, the calculated $(\alpha,\gamma)$ and $(\alpha,n)$ cross sections shown in Fig.~\ref{Fig:Cd108ax}(a) and Fig.~\ref{Fig:Cd108ax}(b), respectively, are obtained using the EGLO parameters for $^{116}$Sn \cite{hkt11} as well as the related SLO and GLO models for $E$1 radiations, along with the SLO one for $M$1 radiation. 
These results indicate that the $\alpha$-particle OMP \cite{va14} and only the EGLO model provide a good agreement with the measured $(\alpha,\gamma)$ excitation function \cite{ps16} and $\Gamma_{\gamma}$ value estimated on the basis of RIPL-3 \cite{ripl3} (also in Table~\ref{densp}). 
Moreover, the related overpredictions by the GLO and especially SLO models go well beyond the uncertainty band corresponding to the estimated limits of the level-density parameter $a$ (Table~\ref{densp}) for $^{112}$Sn. 
Obviously, the RSF effects on the $(\alpha,n)$ cross sections shown in Fig.~\ref{Fig:Cd108ax}(b) are within the measured-data errors.

On the other hand, we found this case useful for checking the RSFs obtained for neighboring nuclei. 
Thus, we used also the EGLO parameters for $^{112}$Cd \cite{acl13} that were provided for a couple of $T_f$-parameter values. 
The change shown in Fig.~\ref{Fig:Cd108ax}(c) is lower than even half of the data-error bars and within or close to the uncertainty band given by the limits of the estimated $\Gamma_{\gamma}$ value (Table~\ref{densp}). 
Therefore the use of the RSF of neighboring nuclei is supported in the case of no measured data for a given nucleus. 
At the same time, one may note the rather similar uncertainty bands related to NLD and RSF parameters in Figs.~\ref{Fig:Cd108ax}(a) and \ref{Fig:Cd108ax}(c), respectively.

A particular remark concerns a notable involvement of the ratio between $(\alpha,\gamma)$ and $(\alpha,n)$ cross sections to remove the sensitivity of the adopted $\alpha$-particle OMP \cite{ps16}. 
However, while this ratio was much more sensitive to the adopted RSF model, the SLO model turned out to best fit the  measured ratios. 
As it was argued \cite{ps16}, this result does not imply that, unlike the particular combination of SM parameters including it, this model is the best. 
Moreover, this outcome at variance with the results of RSF data analysis underlines the advantage of using a consistent parameter set established by means of various independent data analysis.

We also considered the $\alpha$-particle OMP of McFadden and Satchler \cite{lmf66} which was used in Ref. \cite{ps16}. 
First, a good agreement is provided by this OMP at the higher energies of the measured $(\alpha,\gamma)$ and  $(\alpha,n)$ cross sections \cite{ps16}. Then, at lower energies there is a small overestimation of the $(\alpha,n)$ but a significant one of the $(\alpha,\gamma)$ data.
Moreover, while this OMP and above-mentioned RSF effects on calculated $(\alpha,n)$ data are small and in reverse order [Fig.~\ref{Fig:Cd108ax}(b)], they are quite larger and both overestimating the $(\alpha,\gamma)$ cross sections [Fig.~\ref{Fig:Cd108ax}(a)]. 
Therefore, the deviation of an $(\alpha,\gamma)$ evaluation using these parameters is obvious.

\begin{figure*} 
\resizebox{1.4\columnwidth}{!}{\includegraphics{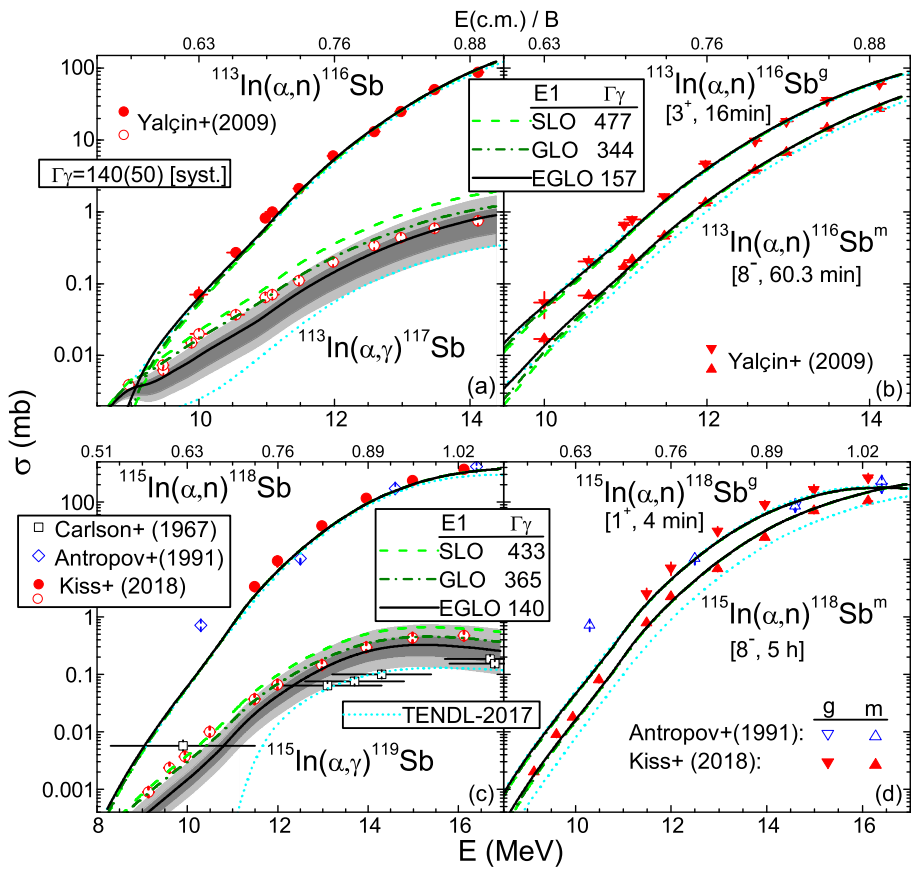}}
\caption{\label{Fig:In113115ax}(Color online)  As  Fig.~\ref{Fig:Zn64ax} but for $^{113,115}$In nuclei \cite{ggk18,exfor,cy09} and (b,d) the g.s. as well as isomeric $(\alpha,n)$ cross sections, except (a,c) the uncertainties corresponding to limits (Table~\ref{densp}) of the level-density parameter $a$ for $^{116,118}$Sb (gray bands), as well as including the ones of $\Gamma_{\gamma}$ value based on data systematics \cite{ripl3} (light-gray bands).}
\end{figure*}

The calculated $(\alpha,n)$ cross sections are naturally close by the TENDL-2017 evaluation [Fig.~\ref{Fig:Cd108ax}(d)] that used the $\alpha$-particle OMP \cite{va14} as the default option in the TALYS-1.9 code. 
However, the large TENDL-2017 deviation of $(\alpha,\gamma)$ evaluation [Fig.~\ref{Fig:Cd108ax}(c)] denotes a quite different case entirely due to the RSF account.

\subsection{$(\alpha,x)$ reactions on $^{113,115}$In} \label{In}

Particular attention should be paid to the conclusion that further efforts are needed to establish an OMP that simultaneously describes $\alpha$-particle elastic scattering \cite{ggk16} and $(\alpha,\gamma)$ and $(\alpha,n)$ reactions on $^{115}$In \cite{ggk18}. 
Part of this conclusion was due to an excellent description \cite{ggk16} of the elastic scattering data with the OMP \cite{va14}, at the same time with a significant underestimation of $(\alpha,\gamma)$ and isomeric $(\alpha,n)$ cross sections at lower energies \cite{ggk18}. 
However, a previous analysis of the $\alpha$-induced reaction data on $^{113,115}$In below and around $B$ (e.g., \cite{cy09}) was carried out \cite{ma10} with a rather good agreement with all data available at that time. 
This is why we found of interest the inclusion of the newest $(\alpha,x)$ reaction data for $^{115}$In within a revision of the former analysis.

An additional aim of this work is to use   the EGLO parameters of the RSF model, that were established more recently for the $^{117}$Sn excited nucleus \cite{hkt11}. 
However, in spite of the corresponding $\Gamma_{\gamma}$ values in rather good agreement [Table~\ref{densp} and Figs.~\ref{Fig:In113115ax}(a,c)] with systematics of the measured data \cite{ripl3}, the $(\alpha,\gamma)$ data for both $^{113,115}$In are underestimated just above the $(\alpha,n)$ threshold. 
This underestimation includes the uncertainties corresponding to the limits (Table~\ref{densp}) of both the CN level-density parameter $a$ and the $\Gamma_{\gamma}$ value based on data systematics \cite{ripl3}. 
The latter limits correspond to large uncertainties which were assumed due to the scarce $\Gamma_{\gamma}$ data \cite{ripl3} available for odd-even excited nuclei.
They were used for an additional RSF normalization that led to the calculated $(\alpha,\gamma)$ cross sections alongside the light-gray uncertainty bands in Fig.~\ref{Fig:In113115ax}.

The GLO and SLO models provide a much better agreement only for reaction data while the related $\Gamma_{\gamma}$ values  are larger than the systematical estimation by a factor $>$2. 
As the former analysis of the RSF data \cite{hkt11} supports only the EGLO model, the questions on these $(\alpha,\gamma)$ excitation functions remain open. Maybe the actual knowledge of the neutron-deficient odd Sb (Z=51) isotopes, with one valence proton, needs further improvement in order to make possible a realistic account of their structure.  

\begin{figure*} 
\resizebox{1.5\columnwidth}{!}{\includegraphics{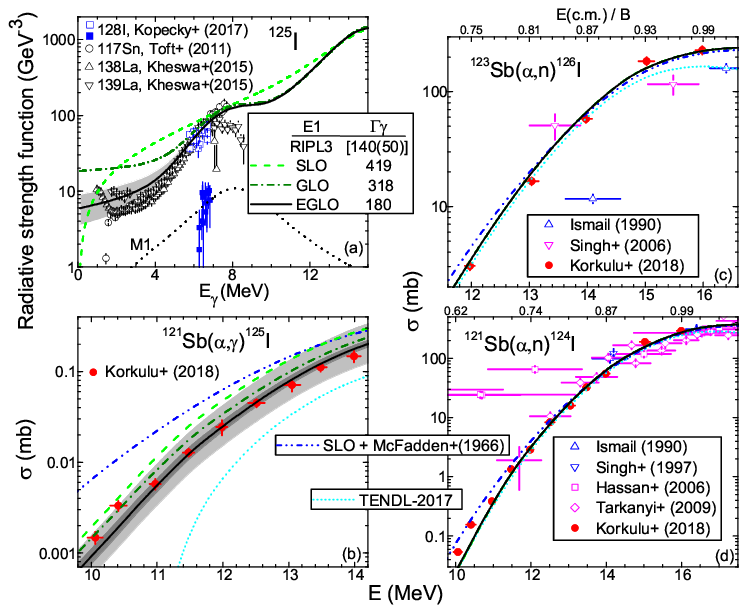}}
\caption{\label{Fig:Sb121123ax}(Color online) (a) Comparison of experimental \cite{hkt11,bvk15,jk17} and calculated sum of $\gamma$-ray strength functions of the $E1$ and $M1$ radiations for $^{125}$I using the models SLO (dashed curve), GLO (dash-dotted curves), and EGLO (solid curves), for $E1$ radiations, and the SLO model for $M1$ radiations (dotted curve), while the RSF uncertainty (light-gray band) corresponds to $T_f$=(0.6$\pm$0.14) MeV.
The $s$-wave average radiation widths $\Gamma_{\gamma}$ (in meV) either deduced from systematics \cite{ripl3}, or corresponding to $M1$ and each of above-mentioned $E1$ functions are also shown.
(b-d) As Fig.~\ref{Fig:Zn64ax} but for $^{121,123}$Sb nuclei \cite{zk18,exfor}, except the calculated data using the SLO model and the $\alpha$-particle OMP of McFadden and Satchler \cite{lmf66} (dash-dot-dotted curves), and (b) uncertainties due to those of RSF one (gray band) and also NLD (light-gray band).}
\end{figure*}

On the other hand, the g.s. as well as isomeric $(\alpha,n)$ cross sections are well described in the limit of either the error bars, for $^{113}$In [Fig.~\ref{Fig:In113115ax}(b)] and half of the data for $^{115}$In [Fig.~\ref{Fig:In113115ax}(d)], or 2$\sigma$ uncertainty for the rest of $^{115}$In data. 
As the sum of these cross sections is an order of magnitude larger than the $(\alpha,\gamma)$ cross sections shortly above the  $(\alpha,n)$ reaction threshold (Fig.~\ref{Fig:In113115ax}), the OMP \cite{va14} is validated also for  $^{113,115}$In target nuclei. 
However, it is noteworthy that even the $(\alpha,\gamma)$ reaction cross section for $^{113}$In at the lowest energy, just below the $(\alpha,n)$ threshold [Fig.~\ref{Fig:In113115ax}(a)], validates the present analysis while the underestimation corresponds to energies where the neutron emission has a sharp increase. 

Finally, we address the statement \cite{ggk18} regarding the $\alpha$-particle OMP of Ref. \cite{va94} as an earlier version of the actual potential \cite{va14},  optimized mainly at higher energies. 
However, this analysis \cite{va94} addressed the $\alpha$-particle emission in neutron-induced reactions up to $E_n$$\sim$10 MeV, i.e. within several MeV above the corresponding reaction thresholds. Therefore, the above-mentioned emission energies were not higher but below and around $B$. 

As a matter of fact, the involvement at energies lower and around $B$ \cite{va14,ma09,ma10} of an $\alpha$-particle OMP obtained by elastic-scattering analysis well above $B$ \cite{ma03} is indeed a problem. 
However, it has already been proved \cite{ma09a,ma10a,ma10} that one should take into account the particular $\alpha$-particle surface absorption below $B$, the changes of the related $\sigma_R$ being shown in Figs. 1-2 of Refs. \cite{ma09a,ma10a} and Figs. 3-5 of Ref. \cite{ma10}. The proper energy dependence of both the surface and volume components of the $\alpha$-particle imaginary potential has finally been considered \cite{va14}, leading to the suitable account of both the reaction data below $B$ (Fig.~\ref{Fig:In113115ax}) and elastic scattering on $^{113,115}$In \cite{ggk16}.

\subsection{$(\alpha,x)$ reactions on $^{121,123}$Sb} \label{Sb}

The first measurement of the $(\alpha,\gamma)$ cross sections on $^{121}$Sb close to the astrophysically relevant energy range  pointed out a strong overestimation by SM calculations \cite{zk18}. Additionally, $(\alpha,n)$ cross sections were obtained for $^{121,123}$Sb at lower energies  compared to the available data, and especially with much higher precision. Therefore, checking the agreement found earlier for these nuclei (Fig. 3 of Ref. \cite{ma10})  becomes a matter of great interest.

First, we paid closer attention to the RSF account. 
As recent RSF data of nearby nuclei exist only for $^{117}$Sn \cite{hkt11} and $^{138,139}$La \cite{bvk15,bvk17}, available data of $^{128}$I \cite{jk17} have also been compared with the calculated RSFs of $^{125}$I shown in Fig.~\ref{Fig:Sb121123ax}(a). 
The EGLO parameters for $^{117}$Sn \cite{hkt11} led to a suitable RSF average trend when the $T_f$ parameter was increased from 0.46 MeV to 0.6 MeV (Table~\ref{densp}). We considered the related $T_f$-difference of 0.14 MeV as an uncertainty estimation of this parameter and found that the corresponding RSF uncertainty band [Fig.~\ref{Fig:Sb121123ax}(a)] covers well even the RSF low-energy upbend of $^{138}$La \cite{bvk17}. Moreover, while the related RSF change, close to zero energy, is up to 52\%, that of the corresponding $\Gamma_{\gamma}$ value is below 27\% and rather well within the limits of the data systematics.

The GLO and SLO models using the same GDR parameters led to larger RSF values for $\gamma$-ray energies below 5-6 MeV while $\Gamma_{\gamma}$ values increased by factors close or even above 2 [Table~\ref{densp} and Figs.~\ref{Fig:Sb121123ax}(a)]. 
Therefore, despite existing scarce RSF and $\Gamma_{\gamma}$ data, we may consider that a reasonable RSF estimation has finally been obtained particularly with reference to either SLO or GLO models.

Consequently, the use of the EGLO model has led to the agreement with the $(\alpha,\gamma)$ cross sections on $^{121}$Sb \cite{zk18} within the small error bars, except for the two data points at the lowest $\alpha$-particle energies [Fig.~\ref{Fig:Sb121123ax}(b)]. 
These points are well described by the larger values obtained using the GLO and SLO, which however overestimate the rest of this excitation function. 
At the same time, the above-mentioned RSF uncertainty band led to an uncertainty band of the calculated $(\alpha,\gamma)$ cross section going from $\sim$ 40\% at the lowest energy to less than 20\% at the highest one [Fig.~\ref{Fig:Sb121123ax}(b)]. 
However, the additional consideration of the uncertainty of the level-density parameter $a$ (Table~\ref{densp}) yields a total uncertainty band three to five times larger. 
This uncertainty estimation covers all measured data as well as the results of using the GLO model, while the SLO results are even larger.
 
On the other hand, the calculated cross sections using the SLO model of RSF and the $\alpha$-particle OMP of McFadden and Satchler \cite{lmf66}, also shown in Fig.~\ref{Fig:Sb121123ax}(b), seem to be rather close to those obtained by Korkulu {\it et al.} (Fig. 6 of Ref. \cite{zk18}). Moreover, these results indicate that the overestimation by a factor of 2-4 of the measured data has been entirely due to the $\alpha$-particle OMP \cite{lmf66} only at lowest energies. 
The disagreement at the higher energies has been caused by the use of the SLO model for the RSF account.

\begin{figure*} 
\resizebox{1.5\columnwidth}{!}{\includegraphics{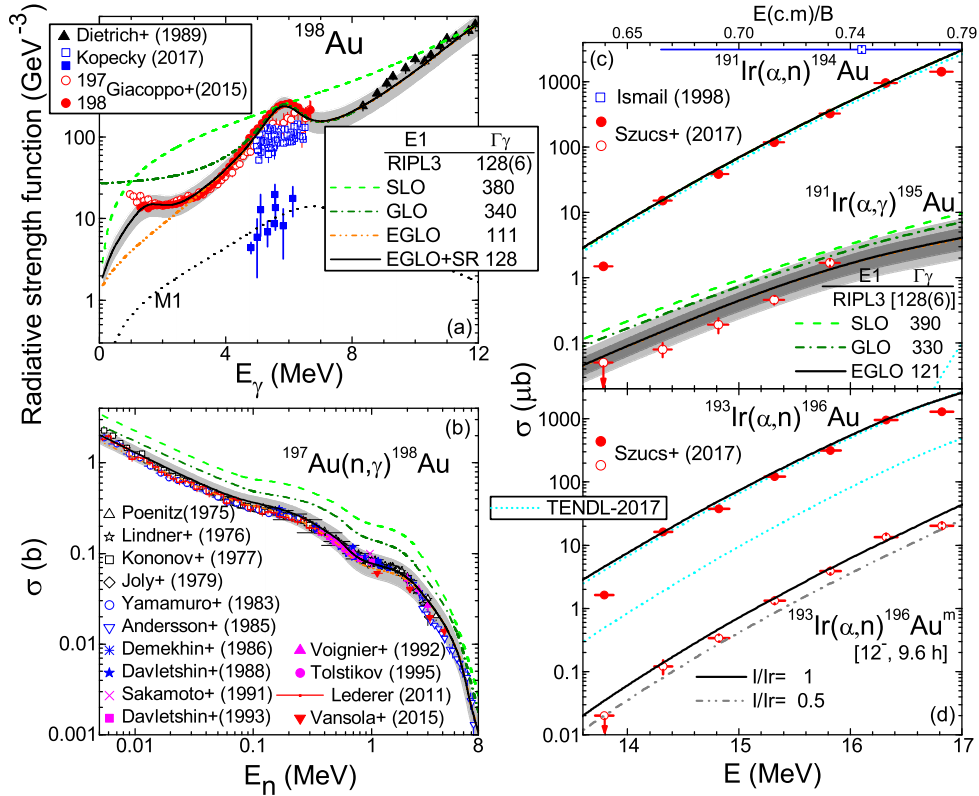}}
\caption{\label{Fig:Ir19193ax} As Fig.~\ref{Fig:Sb121123ax} but for $^{198}$Au \cite{exfor,jk17,fg15} and $^{191,193}$Ir nuclei \cite{ts18,exfor}, except inclusion of (a,b,c) EGLO (dash-dot-dotted curves) and EGLO+SR (solid  curves), (b) comparison of neutron-capture cross sections for $^{197}$Au, measured \cite{exfor} and calculated using the same $E1$ radiation RSF models but only the uncertainty band (light gray) due to that of RSF, and (d) additional calculated cross sections corresponding to $I$/$I_r$=0.5 (dash-dot-dotted curve) }
\end{figure*}

The comparison of the measured and calculated $(\alpha,n)$ cross sections for $^{121,123}$Sb  [Figs.~\ref{Fig:Sb121123ax}(c,d)] can take the advantage of the recent data unprecedented precision. 
While one could not differentiate between $\alpha$-particle OMPs \cite{ma10,lmf66} by means of the data measured even in the last decade or so, there is now a new case  particularly for $^{123}$Sb [Fig.~\ref{Fig:Sb121123ax}(c)]. 
A slight difference between the predictions of the two OMPs \cite{va14,lmf66}, much smaller than the error bars of previous measurements, supports now the potential \cite{va14}. 
On the other hand, the use of the same potential as the default option of TALYS-1.9 led to TENDL-2017 evaluated $(\alpha,n)$ data close to the present calculation at energies where the nuclear level densities and PE effects are not yet playing a significant role.

The calculated and measured data of the $(\alpha,n)$ reaction on $^{121}$Sb  [Fig.~\ref{Fig:Sb121123ax}(d)] are in a very similar situation to that of $(\alpha,\gamma)$ excitation function. 
Only the calculated values at the two lowest energies are underestimating the experimental ones \cite{zk18}, yet below a 2$\sigma$ uncertainty. 
Nevertheless, the case of $^{121}$Sb target nucleus is one of the very few in which the analysis of both $(\alpha,n)$ and $(\alpha,\gamma)$ reactions is necessary to validate an $\alpha$-particle OMP, provided that the involved RSF has also been proved in advance.

\subsection{$(\alpha,x)$ reactions on $^{191,193}$Ir} \label{Ir}

The similar measurement with a previously unprecedented sensitivity on $^{191,193}$Ir by Sz\"ucs {\it et al.} \cite{ts18} enables an extension of the present work for heavy nuclei. 
Moreover, a recent measurement of RSF data for $^{198}$Au  as well as its involvement within the well-known standard $^{197}$Au$(n,\gamma)^{198}$Au reaction analysis \cite{fg15} provide better conditions for $(\alpha,\gamma)$ reaction suitable account. 

Thus, we adopted an RSF energy dependence rather similar to Figs. 5-6 of Ref. \cite{fg15} using the EGLO model and (i) the SLO parameters of both $E1$ and $M1$ radiations of Kopecky-Uhl \cite{jk90}, (ii) the EGLO parameter $T_f$ of Giacoppo {\it et al.} \cite{fg15} as well as (iii) their pigmy dipole resonance (PDR) parameters with the PDR cross section of 12.2 mb, and (iv) low-energy small resonance (SR) tail (model A in Table II of Ref. \cite{fg15}). 
First, there are less significant SR effects on the suitable EGLO calculated values of RSF [Fig.~\ref{Fig:Ir19193ax}(a)] and $\Gamma_{\gamma}$ (Table~\ref{densp}).
Second, the GLO and SLO models led to either RSFs well beyond an uncertainty band corresponding to the average change of 30\% (shown in Fig. 3(c) of Ref. \cite{fg15} to follow the RSF normalization using various spin-distribution models), or $\Gamma_{\gamma}$ values increased by an average factor of 3. 
Actually, additional effects due to the minor error bars of $D_0$ and $\Gamma_{\gamma}$ experimental values \cite{ripl3} (Table~\ref{densp}) were not considered anymore because the use of NLD parameters related to the limits of the fitted $D_0^{\it exp}$ provides changes of $\Gamma_{\gamma}$ well within its own error bar. 

The EGLO model corresponds also to calculated $^{197}$Au$(n,\gamma)^{198}$Au reaction cross sections [Fig.~\ref{Fig:Ir19193ax}(b)] in good agreement with the measured values \cite{exfor} that are nearly all within the uncertainty band corresponding to the above-mentioned one for the adopted RSF.
The SR contribution is noticeably improving this agreement only at the lowest $\gamma$-ray energies, inside the same uncertainty band.
The same excitation function is substantially overestimated by the GLO and SLO models using the same GDR parameters  by factors of $\sim$2 and $\sim$3, respectively, even without the SR addition.

The calculated $(\alpha,\gamma)$ excitation function for $^{191}$Ir, using the EGLO model for the electric-dipole RSF, is in agreement with the measured data in the limit of 2$\sigma$ uncertainty [Fig.~\ref{Fig:Ir19193ax}(c)]. 
Moreover, these data are within the uncertainty bands corresponding to the limits of the level-density parameter $a$ for the compound nucleus $^{195}$Au (Table~\ref{densp}) and, in addition, an RSF systematic uncertainty of 30\% similar to that for $^{198}$Au \cite{fg15}. 
The GLO and SLO models are leading to larger $(\alpha,\gamma)$ cross sections and $\Gamma_{\gamma}$ values by factors of over 2 and around 3 (also in Table~\ref{densp}]), respectively. 
On the other hand, it seems that the experimental excitation functions has a faster slope than predicted by our calculations or Ref. \cite{ts18}. 
However, the check of the adopted RSF described above, including the suitable account of the $^{197}$Au$(n,\gamma)^{198}$Au cross sections, provide confidence in the $\gamma$-ray and neutron competition we assume. 

The comparison between the measured  $(\alpha,n)$ cross sections for $^{191,193}$Ir and calculated results obtained with the optical potential \cite{va14} shows a good agreement except the data points measured at the lowest and highest incident energies [Figs.~\ref{Fig:Ir19193ax}(c,d)].  
Nevertheless, a continuous increase with energy is shown not only by TENDL-2017 evaluation, which was obtained with the same OMP \cite{va14}, but also by Sz\"ucs {\it et al.} \cite{ts18} with a modified version of McFadden and Satchler potential \cite{lmf66}. 
The change of the latter OMP  consists in the replacement of the volume imaginary-potential constant depth $W$=25 MeV with a Fermi-type function at an energy 0.9$B$ \cite{ma10} and having a 'diffuseness' $a_E$ used as a free parameter. 
Sz\"ucs {\it et al.} found the best description of their $(\alpha,n)$ experimental data \cite{ts18} using a
value $a_E$=(2$\pm$0.5) MeV corresponding to the limits of the data except that at the highest energy. 
Our overestimation is rather similar at the lowest energy to that of their best fit, but lower by a factor of 3--4 at the highest energy. 

The overestimation of experimental $(\alpha,\gamma)$ cross sections for $^{191}$Ir even by the modified OMP corresponding to the lower parameter value $a_E$=1.5 MeV \cite{ts18} is also notable.
On the other hand, different values $a_E$=4--6 MeV were found earlier to provide an excellent reproduction of the experimental cross sections of $^{187}$Re$(\alpha,n)^{190}$Ir reaction \cite{ps14}. 
However, with no further change, the OMP \cite{va14} provides a similar description of the data for $^{187}$Re \cite{va16} as well as an improved one for $^{191,193}$Ir.

Moreover, we obtained an even better description of the isomeric cross sections of the  $^{193}$Ir$(\alpha,n)$ reaction in comparison with the related total cross sections [Fig.~\ref{Fig:Ir19193ax}(d)]. 
Actually, the high-spin second isomeric state of $^{196}$Au is the 55th excited state of the residual nucleus at the top of the discrete levels taken into account in SM calculations (Table I of Ref. \cite{ma12}). 
Therefore its population follows the side feeding and continuum decay, being particularly determined by the $\alpha$-particle OMP, nuclear level density, and RSF. 
While proved the appropriate assumptions for the latter quantities \cite{ma12b}, the effective nuclear moment of inertia $I$ which is most important for the isomeric cross section estimation may still be uncertain. 
Thus, although neutron-induced data analysis suggested a constant $I_r$ value for the effective $I$ of $^{198}$Au \cite{ma12}, more recent RSF and $(n,\gamma)$ reaction data suggest that levels in the quasicontinuum are dominated by lower spins \cite{fg15}. 

Consequently the isomeric cross sections of the  $^{193}$Ir$(\alpha,n)$ reaction were  calculated by using a constant value 0.5$I_r$. 
The corresponding results also shown in Figs.~\ref{Fig:Ir19193ax}(d) prove, however, a lower sensitivity of the calculated $(\alpha,n)$ isomeric cross sections to this quantity, than the neutron activation data (e.g., Fig. 1 \cite{ma12}). 
Thus, using a 0.5$I_r$ value led to underestimated cross sections within the limit of 2$\sigma$ uncertainty.
Nevertheless, the agreement of the measured and calculated isomeric data is noteworthy as long as the related TENDL-2017 evaluation, which reproduces the measured total $(\alpha,n)$ excitation function, shows more than one order of magnitude larger isomeric cross sections.

\section{SUMMARY AND CONCLUSIONS} \label{Conc}

Our analysis of recent high-precision measurements of $\alpha$-induced reactions on $^{64}$Zn, $^{108}$Cd, $^{113,115}$In, $^{121,123}$Sb, and $^{191,193}$Ir, below the Coulomb barrier, points out eventual uncertainties and/or systematic errors of an $\alpha$-particle OMP \cite{va14} assessment as follows. 
In any case, independent data were used in advance to establish or validate uncertain statistical-model parameters. 
Moreover, we took notice of the calculated cross-section uncertainties related to the error-bar limits of level-density parameters and $\gamma$-ray strength functions, that follow the above-mentioned independent data accuracy limits (Table~\ref{densp}). 
A distinct case is that of the zero-energy upbend of the $M1$-radiation RSF, which was taken into account only for $^{68}$Ge nucleus (Figs.~\ref{Fig:RSF_GeF} and \ref{Fig:Zn64ax}), while elsewhere we adopted the usual SLO parameters used within former analyses \cite{hkt11,jk90,acl13,bvk15,bvk17,fg15}.

(i) Consistent SM parameters alongside the $\alpha$-particle potential \cite{va14} can provide a reliable account of all available data \cite{ao16,gg12} and $\chi^2$-based  predictions \cite{pm17} of $\alpha$-induced reactions on $^{64}$Zn. 
The uncertainty bands corresponding to the adopted RSF and NDL parameter uncertainties cover the recently measured $(\alpha,\gamma)$ cross sections, while the results following the alternative use of SLO and GLO models of electric-dipole RSFs are well above them.
The deviations within 2$\sigma$ uncertainty of the measured data at several incident energies, particularly for the  $(\alpha,n)$ reaction, are supported by the former trial of independent data.

(ii) The precise cross sections of the $(\alpha,\gamma)$ and $(\alpha,n)$ reactions on $^{108}$Cd \cite{ps16} prove not only the $\alpha$-potential \cite{va14} but also the use of RSFs of neighboring nuclei for a given nucleus with no similar data. 
On the other hand, the notable involvement of the ratio between $(\alpha,\gamma)$ and $(\alpha,n)$ cross sections to remove the sensitivity of the adopted $\alpha$-particle OMP \cite{ps16}, may lead to results at variance with the primary RSF data analysis.

(iii) The total $(\alpha,n)$ as well as g.s. and isomeric $(\alpha,n)$ cross sections for $^{113,115}$In  \cite{ggk18} have also been well described. 
However, despite of the rather good agreement of the EGLO values with the systematics of measured $\Gamma_{\gamma}$ \cite{ripl3}, the $(\alpha,\gamma)$ reaction cross sections of both $^{113,115}$In are underestimated above the $(\alpha,n)$ threshold. 
The GLO and SLO models provide a much better agreement for reaction data but  large overestimate $\Gamma_{\gamma}$. 
Nevertheless, the $(\alpha,\gamma)$ reaction cross section for $^{113}$In at the lowest energy just below the $(\alpha,n)$ threshold validates, however, this potential while the underestimation corresponds to energies where the $\gamma$ channel weakens. 
Actually, the suitable account of the reaction data below $B$ \cite{ggk18} as well as elastic scattering on $^{113,115}$In \cite{ggk16} do support the proper energy dependence of both surface and volume components of the $\alpha$-particle imaginary potential \cite{va14}. 
Maybe the actual knowledge of the neutron-deficient odd Sb (Z=51) isotopes may need further improvement in order to enable a realistic account of their structure also involved in SM calculations.  

(iv) The reasonable estimation of the RSF of $^{128}$I by the EGLO model, especially with reference to either SLO or GLO models, was accompanied by a general agreement with the $(\alpha,\gamma)$ cross sections on $^{121}$Sb \cite{zk18}. 
Moreover, the use of the $\alpha$-particle OMP of McFadden and Satchler \cite{lmf66} and the SLO/GLO models for RSF has small effects on calculated $(\alpha,n)$ cross sections and even in reverse order, e.g. for $^{108}$Cd, but quite larger and both overestimating the $(\alpha,\gamma)$ data. 
Moreover, the overestimation by a factor of 2-4 of the measured data has been entirely due to the $\alpha$-particle OMP \cite{lmf66} only at lowest energies. 
The disagreement at the higher energies has been caused by the use of the SLO model for the RSF account.

(v) The EGLO model for the electric-dipole RSF, formerly proved including the suitable account of the $^{197}$Au$(n,\gamma)^{198}$Au cross sections,  supports an increase among the measured data in the limit of 2$\sigma$ uncertainty of the calculated $(\alpha,\gamma)$ excitation function for $^{191}$Ir. 
The comparison between the measured $(\alpha,n)$ cross sections for $^{191,193}$Ir and calculations using the optical potential \cite{va14} shows a good agreement except the data points measured at the lowest and highest incident energies.  
Moreover, we obtained a better account for the isomeric cross section of the $^{193}$Ir$(\alpha,n)$ reaction compared to the related total cross section. 

A final remark concerns assumption \cite{ggk18} of the $\alpha$-particle OMP of Ref. \cite{va94} as an earlier version of the actual potential \cite{va14},  optimized mainly at higher energies. 
However, this analysis \cite{va94} addressed the $\alpha$-particle emission in neutron-induced reactions up to $E_n$$\sim$10 MeV, i.e. within several MeV above the corresponding reaction thresholds. Therefore, the above-mentioned emission energies were not higher but below and around $B$. 

At the same time, the involvement at energies lower and around $B$ \cite{va14,ma09,ma10} of an $\alpha$-particle OMP obtained by elastic-scattering analysis well above $B$ \cite{ma03} should take into account the particular $\alpha$-particle surface absorption below $B$. 
Thus, changes of the related $\sigma_R$ were shown in Figs. 1-2 of Refs. \cite{ma09a,ma10a} and Figs. 3-5 of Ref. \cite{ma10}. 
That said, changes of both surface and volume imaginary potentials correspond, through the dispersive relations with an integral over all incident energies, to a change of, e.g., the semi-microscopic real potential. 
Former $\alpha$-particle elastic-scattering analyses \cite{ma03,ma09} have shown lower sensitivity to the addition of the dispersive correction to DFM real potential.
A deeper insight may follow further precise measurements including cross sections of major reaction channels that may not be well described at the moment.

Finally, it seems that the alpha-nucleus OMP \cite{va14}, even though far from perfect, looks like a reasonable compromise when adopted in the extensive model calculations of present interest.
Last but not least, the use of the same potential \cite{va14} as the default option of TALYS-1.9 led to TENDL-2017 evaluated $(\alpha,n)$ data close to our calculation at energies where the nuclear level densities and PE effects are not yet playing a significant role. However, the large TENDL-2017 deviation for $(\alpha,\gamma)$ reactions and the isomeric cross sections of $^{193}$Ir$(\alpha,n)$ reaction highlights the importance of a suitable account of all reaction channels for $\alpha$-nucleus optical potential validation.

\section*{Acknowledgments}

The authors acknowledge the useful discussions and correspondence with Jura Kopecky.
This work has been partly supported by Autoritatea Nationala pentru Cercetare Stiintifica (Project PN-18090102), within the framework of the EUROfusion Consortium and has received funding from the Euratom research and training program 2014-2018 and 2019-2020 under Grant Agreement No. 633053. The views and opinions expressed herein do not necessarily reflect those of the European Commission.

\bibliography{CX10563rev3}

\end{document}